\documentclass[prc,aps,twocolumn,psfig,amsmath,amssymb]{revtex4}
\usepackage{graphicx}\usepackage{epsfig}
\usepackage{amsmath}\usepackage{amssymb}\usepackage{amsfonts}
\usepackage{bm}
\usepackage{exscale}
\usepackage{dcolumn}    
\usepackage{color}
\def\new#1{{#1}}
\def\diff{\mathrm{d}}
\def\bred{b_{\mathrm{red}}}
\newcommand{\Etr}{E_{\textrm{tr}}}
\newcommand{\Eperp}{E_{\perp}}
\newcommand{\egeo}{e_{\textrm{geo}}}
\newcommand{\phiplane}{\phi_{\textrm{plane}}}
\newcommand{\phiAM}{\phi_{\textrm{AM}}}

\begin{document}
%
%
\title{
	Probing the nuclear equation of state 
in heavy-ion collisions at Fermi energy \\
in isospin-sensitive exclusive experiments
}
%
%
\author{P.Napolitani$^{1,2}$}
\author{M.Colonna$^{3}$}
\author{F.Gulminelli$^{1}$}
\author{E.Galichet$^{2,5}$}
\author{S.Piantelli$^{4}$}
\author{G.Verde$^{3}$}
\author{E.Vient$^{1}$}
%
%
\affiliation{$^{1}$~LPC Caen, ENSICAEN, Universit\'e de Caen, CNRS/IN2P3, 14050 Caen cedex 4, France}
\affiliation{$^{2}$~IPN, CNRS/IN2P3, Universit\'e Paris-Sud 11, 91406 Orsay cedex, France}
\affiliation{$^{3}$~INFN-LNS, Laboratori Nazionali del Sud, 95123 Catania, Italy}
\affiliation{$^{4}$~Sezione INFN di Firenze, Via G. Sansone 1, 50019 Sesto Fiorentino, Italy}\affiliation{$^{5}$~Conservatoire National des Arts et M\'etiers, 75141 Paris Cedex 03, France}
%
%
\begin{abstract}
\begin{center}
	For the FAZIA collaboration
\end{center}
%
\new{
	We propose a few selected experimental approaches 
to show that new-generation instruments can give a direct access to 
significant observables on the density dependence of the symmetry 
energy in the nuclear equation of state.
	The form of such dependence is investigated within the 
Stochastic Mean Field model, coupled to a secondary-decay treatment.
}
\end{abstract}
\maketitle
%
%
%
%
\section{
    Introduction				\label{section1}
}
%
%
	The exotic-beam facilities which already exist, or 
which are under construction, invite to focus on
a longstanding experimental challenge: heavy-ion collisions induced 
by exotic nuclei at low and intermediate energies 
will be a probe for
the properties of isospin-asymmetric nuclear matter~\cite{LoI}.
	In particular, numerous next-generation experiments will 
be dedicated to the study of the density behaviour of the symmetry 
energy. 

	Such quantity defines the isospin-asymmetric part of the 
equation of state (EOS), which can be 
deduced from different phenomenological 
parametrisations~\cite{Fuchs2006,Li2008,Baran2005}.
	The corresponding predictions can be grouped schematically
into two main forms: either a `stiff' or a `soft' dependence of 
the symmetry energy as a function of the nuclear 
density~\cite{Baran2005,Shetty2007}.
	At present, considerable effort is dedicated to pin down 
the density dependence in the regime of relativistic heavy-ion 
reactions~\cite{Li2008,Xiao2009,Baran2005}, where supersaturation 
densities are accessible.
	Such conditions determine a significant difference between 
the two forms for the density dependence but
the uncertainties in the hadron effective mass splittings and momentum 
dependence in the isovector channel 
introduce new degrees of freedom, enhancing the complexity of the problem
~\cite{Ditoro2009}.

	These complications do not arise in the Fermi-energy 
regime, where the functional form of the symmetry energy at 
subsaturation densities should in principle be accessible.
	Furthermore, these conditions give access to the 
interesting issue of the influence of cluster 
correlations~\cite{Horowitz2006,Lehaut2009}. 
	At Fermi energies, the difficulty is that the difference 
between the theoretical predictions for the density dependence is 
less pronounced than at 
the large supersaturation densities which can be probed at 
relativistic energies:
in particular,
it is not always clear whether an effective discrimination between 
different degrees of asy-stiffness is really possible.
	On the one hand, this is due to the dependence of the
predictions on the existing transport 
models~\cite{Li2008,Ditoro2009,Tsang2009}. 
	On the other hand, besides the detection limitations, also
the secondary decay contributes in deforming EOS-sensitive isotopic 
observables~\cite{Liu2004,Shetty2006,Tian2008}: we will discuss how
this can affect the comparison protocol between model and experiment.

	In this respect, we need to identify isotopic observables which,
first of all, manifest a significant sensitivity to the change 
between the different asy-EOS forms and, in addition, constitute 
robust experimental observables against the effect of secondary 
decay and detection limitations~\cite{Colonna2006,Ditoro2006,Baran2009,Tsang2009}, is necessary.
\new{	We also indicate that, recently, such study has profited 
from the results of isospin-transport experiments, which already 
imposed theoretical constraints~\cite{Tsang2004,Chen2005,Shetty2007,Galichet2009,DeFilippo2009,Amorini2009}.
}\new{
	This work is aimed to give a schematic guideline for 
measuring a selection of isospin observables with new-generation
isospin-sensitive instruments, in comparison with former isospin-blind 
devices.
	In particular, within our simulation protocol, we focus on
a forthcoming $4\pi$ detector, \emph{FAZIA}~\cite{FAZIAweb}, 
explicitly planned for measuring isospin observables and its 
isospin-blind ancestor, \emph{INDRA}~\cite{Pouthas1995,Pouthas1996}.
	We only focus on these two detectors because they are well 
suited for representing two opposite situations.
	However, it is evident that they can not resume the several 
experimental strategies which exist or are planned and which 
are based on different innovative tools for accessing the isospin 
observables.
	In particular, the $4\pi$ detector \emph{CHIMERA}~\cite{Pagano2004} 
is a new-generation detector which is rapidly improving and 
evolving between these two extremes and its experimental results 
already mark the way towards future isospin-sensitive devices.
}
\section{
    Description of the model				\label{section1bis}
}
	In this paper, a collision system is described within the 
Stochastic Mean Field (SMF) model.
	This is a time-dependent semi-classical mean-field model 
where nucleon-nucleon collisions as well as fluctuations are taken 
into account in the Boltzmann-Uehling-Uhlenbeck formalism 
(see \cite{Rizzo2008_blob} and references therein); the references 
\cite{Baran2005,Rizzo2008} give theoretical and numerical details.
	Such model is reliable in the Fermi-energy regime, for an 
incident energy ranging between about 10 and 200 MeV per nucleon.
	In particular, due to the realistic isospin-dependent mean-field 
and the introduction of density fluctuations, the model gives a reliable 
description of heavy and intermediate-mass fragment properties;
since elastic collision processes are accounted for, the model also 
describes the production of preequilibrium neutrons and protons
reliably; nevertheless, the description of light particles demands 
correlations which are out of the scope of the model.
 
%
%
\begin{figure}[t]\begin{center}
\includegraphics[angle=0, width=0.82\columnwidth]{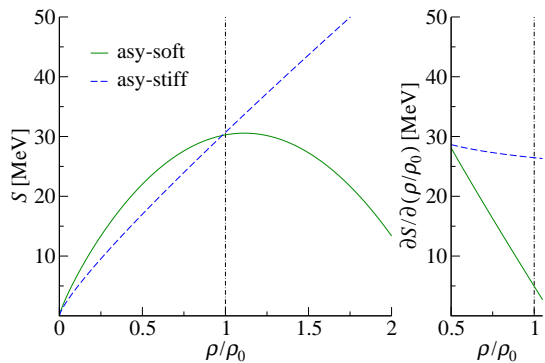}
\end{center}\caption
{
	(Color online)
	\new{Left. Asy-soft and asy-stiff forms for the symmetry energy.
	Right. Corresponding derivative with respect to the density
	in the low-density region}
}
\label{fig1}
\end{figure}

\new{
	In the calculations, we adopt a parametrisation which
gives the same properties as the SkM*\cite{Bartel1982} force for 
nuclear matter: these are a saturation density 
$\rho_0=0.16$~fm$^{-3}$ and an incompressibility modulus
$K_{\infty} = 200\,$MeV which corresponds to a soft equation of 
state.
	We adopt two different prescriptions for the behaviour of 
the symmetry energy $S$, which are respectively:
\begin{eqnarray}
S_{\textrm{stiff}}&=&a\left(\frac{\rho}{\rho_0}\right)^{2/3}+\frac{b}{2}\frac{\rho}{\rho_0}\, ,
\\
S_{\textrm{soft}}&=&a\left(\frac{\rho}{\rho_0}\right)^{2/3}+\frac{1}{2}\left(c\rho+d\rho^2\right)\, , 
\end{eqnarray}
with the parameters
$a=12.7\,$MeV,
$b=36\,$MeV,
$c=481.7\,$MeV~fm$^3$, and
$d=-1638.2\,$MeV~fm$^6$.
	As illustrated in fig.~\ref{fig1},
the first is an `asy-stiff' form, for which the potential 
symmetry term linearly increases with nuclear density, while the 
second is an `asy-soft' form, corresponding to a flatter behaviour 
of the potential symmetry energy around and below normal density.
More details are provided in Ref.\cite{Baran2005}.
}

%
%
\begin{figure}[b]\begin{center}
\includegraphics[angle=0, width=1.\columnwidth]{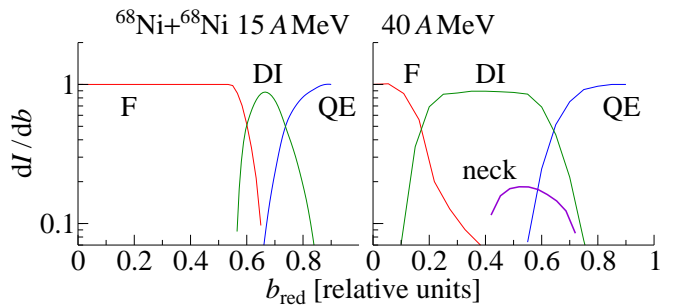}
\end{center}\caption
{
	(Color online)
	Probability for different reaction mechanisms as a function of
the reduced impact parameter for the systems $^{68}$Ni$+^{68}$Ni
at 15 (left panel) and 40 (right panel) $A$MeV.
	The asy-soft (here used) and asy-stiff (not shown) forms give 
very close results.
}
\label{fig2}
\end{figure}
Isospin effects originate from the fact that neutrons and protons experience different
forces. In particular,  the difference between the
neutron and proton currents $\boldsymbol{j}_n -\boldsymbol{j}_p$, that 
develop in presence of asymmetry ($\nabla I$)  and/or density  ($\nabla \rho$)
gradients, is stricly connected to the strength of the symmetry energy (and of its derivative). 
In fact, within a simple hydrodynamical picture, one can write:
\begin{equation}
	\boldsymbol{j}_n -\boldsymbol{j}_p \propto
		\underbrace{S(\rho)\nabla I}_{\textrm{diffusion}} +
		\underbrace{\frac{\partial S(\rho)}{\partial\rho}I\nabla\rho}_{\textrm{migration}}
\;.
\label{eq_3}
\end{equation}
Hence, in presence of asymmetry gradients (diffusive processes) we test essentially the strength of the symmetry energy 
while, when density gradients are encountered along the dynamical path, we observe ``isospin migration''
towards the low density regions, ruled by the derivative of the symmetry energy. 

	Within the SMF model, the systems 
$^{68}$Ni$+^{68}$Ni, $^{58}$Ni$+^{68}$Ni and $^{58}$Ni$+^{58}$Ni 
at 15 and 40 MeV per nucleon were simulated in the present work for
a continuous distribution of impact parameters $b$, which evolves
as $b \diff b$.

	Fig.~\ref{fig2} shows the reaction mechanisms observed
as a function of the reduced impact parameter 
$\bred$ (normalised to the sum of the nuclear radii of the target
and projectile nuclei). 
%
%
	Fusion reactions (F) were identified 
from the presence at the asymptotic time of a unique fragment having 
bigger mass than the projectile (or the target, equivalently); 
the quasi-elastic channel is defined as non-fusion events where the 
sum of the charges of the two largest fragments exceeds a 
large fraction of the charge of the projectile (or the target, 
equivalently; this fraction is adjusted to 90\% and 85\%, for the 
incident energies of 15 and 40 $A$MeV, respectively); the neck 
contribution (neck) is recognised from events which do not belong 
to the previous categories and where at least three fragments have 
larger charge than helium; all the other events are interpreted as 
dissipative binary reactions (DI).
	This figure shows that a centrality selection corresponds only 
approximately to a selection of the reaction mechanism. 
	This is especially true for the neck events, which never 
dominate the total cross section. 
	Since the experimental apparatus does not deliver the same 
response to the different reaction mechanisms, this may result in 
an uncontrolled bias if unfiltered simulations are confronted to 
the experimental data. 

\subsection{Evolution of the isospin content and 
	effect of the secondary decay}
	In the calculations, the dynamical stage is followed till the 
time t = 260 fm/c. 
	In order to deal with the full range of impact parameters which
are simulated, the fragment properties are evaluated as soon as the 
system breaks up into pieces within this interval of time, in each 
event.
	This procedure implies that events with only one fragment (incomplete
fusion) are followed until the final (longer) time, leading to a reduction
of the calculated  size and excitation energy of the compound system, due
to nucleon emission.
	The output of the SMF calculation is then coupled to the evaporation model 
GEMINI~\cite{Charity1988} (the 2003 release is used). 
	This coupling is necessary because of two reasons.
	First, secondary decay extends over a much longer time than 
the numerical stable interval of time of any transport code; 
Second, even molecular-dynamics models, which in principle can 
describe the production of complex light particles, can not 
reproduce the statistical branching ratios of compound-nucleus 
emission~\cite{Ono2006}.
%
%
\begin{figure}[b]\begin{center}
\includegraphics[angle=0, width=1\columnwidth]{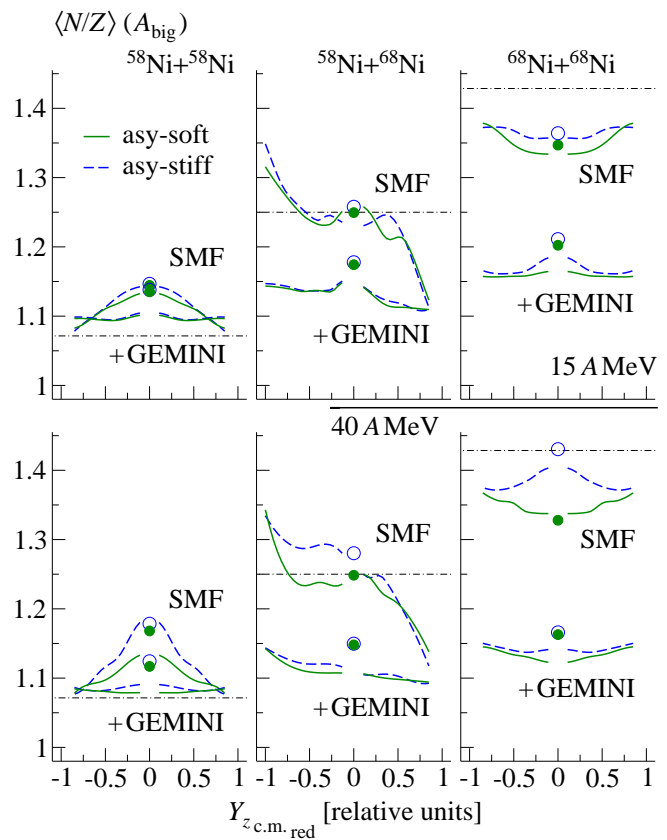}
\end{center}\caption
{
	(Color online)
	Variation of the isotopic composition of the biggest fragment (excluding
 fusion events) as a function of the 
rapidity in the centre of mass  
relative to the projectile (target) rapidity for positive (negative) rapidities 
calculated with both the asy-stiff and asy-soft 
parametrisations before (SMF, upper spectra) and after 
(SMF$+$GEMINI, lower spectra) secondary decay 
for the systems $^{58}$Ni$+^{58}$Ni $^{58}$Ni$+^{68}$Ni and 
$^{68}$Ni$+^{68}$Ni at 15 $A$MeV (upper panels) and 40 $A$MeV (lower panels), for all impact parameters.
	The dots indicate fusion nuclei.
	The dashed lines indicate the average isotopic composition of the
colliding system.
}
\label{fig3}
\end{figure}

	We shall emphasise that in the absence of secondary decay, a basic 
quantity like the isotopic composition of intermediate-mass 
fragments (IMF) would be sufficient to distinguish between the 
asy-stiff and asy-soft behaviour of the symmetry energy, even in 
stable systems.
	This is shown in fig.~\ref{fig3} (upper spectra labelled `SMF'),
where the average neutron number to atomic number ratio of the biggest fragment in all events not
issued of fusion is plotted as a function of the 
corresponding longitudinal component of the rapidity 
in the centre of mass, relative to the projectile (target) rapidity for positive (negative) rapidities, 
for the two systems $^{58}$Ni$+^{58}$Ni, $^{58}$Ni$+^{68}$Ni and 
$^{68}$Ni$+^{68}$Ni at 15 and 40 $A$MeV. 
	In the symmetric systems, the fragments are more neutron rich
in the asy-stiff case and the effect enhances for the more neutron-rich
system. 
\new{
	This scenario can be attributed to the lower value of the 
symmetry energy below normal density in the asy-stiff case, 
with respect to the asy-soft form (see fig.~\ref{fig1}).  
	In the asymmetric system, in addition to nucleon emission,  
isospin diffusion takes place through the low density
interface between the two reaction partners. 
As illustrated by eq.\ref{eq_3},
	diffusion is a process of isospin equilibration between the two 
asymmetric reaction partners which acts more effectively for higher
values of the symmetry energy; at low densities, this process is 
therefore more effective for a softer form of the EOS, which is in 
this case given by the asy-soft parametrisation (see fig.~\ref{fig1}).
	As a consequence, the projectile and target sides in the diagram
of fig.~\ref{fig2} are more similar in this  case, and the 
difference between the two parametrisations is particularly 
relevant at midrapidity, for the system at $40\;A$~MeV
}

	If the reaction ended at this stage in the laboratory,
there would be no need to introduce complex differential 
observables~\cite{Tsang2004}, or to exploit exotic radioactive 
beams for the purpose of this investigation.

	Nevertheless, unfortunately, the secondary decay makes the 
discrimination between the two parametrisations difficult: in stable 
and moderately exotic systems the secondary decay can wash out most of
the discriminating signals; this effect is illustrated in 
fig.~\ref{fig3} (lower spectra labelled `SMF$+$GEMINI'), where the hot 
system described within the SMF model is made decay by the GEMINI 
model.

	We can not exclude that such a prominent effect may depend 
on the model or the system chosen; different contributions of 
secondary decay have been reported in the 
literature~\cite{Colonna2006,Liu2004,Shetty2006,Tian2008,Galichet2009} 
and no systematic study has been performed to our knowledge.
	However, this prologue should drive the attention on the 
fact that the direct measurement of observables related to the 
equation of state is certainly challenging.
	In particular, the evaporation process has the effect of
modifying and even smearing out the observables we wish to
investigate.
	Beside achieving an increasingly more precise understanding 
of the decay sequence (especially in exotic systems), we should also
define efficient strategies in choosing the nuclear system, adopt new observables
and refine the criteria of event selection to reduce the effect of the 
secondary decay, so as to preserve significant signatures.
	These requirements also impose to develop innovative 
experimental strategies.

\subsection{Detection}
	In the present report we suggest some possible solutions 
which would require the use of a dedicated detector device.
	Such detector, which is under development within the
\emph{FAZIA} project (`Four-$\pi$ $A$ and $Z$ Identification 
Array'~\cite{FAZIAweb}), is a $4\pi$-array of telescopes, each one 
designed for measuring the kinetic energy, the nuclear charge and 
the mass of the intercepted fragments (the mass measurement over an extended mass range
constitutes the main innovation for such device).
	An additional calculation filters the results produced by 
the reaction model (SMF+GEMINI) by accurately simulating the 
functioning of the \emph{FAZIA} detector, including the detailed 
geometry and the response of the different detection 
modules~\cite{Napolitani2009}.
	In particular, in the case of \emph{FAZIA}, the telescope 
is composed by two silicon detectors of $300\,\mu$m and $500\,\mu$m 
respectively and a CsI detector; if a particle is arrested in the 
second silicon detector, the kinetic energy and the time of flight 
are measured.
	In this case, the particle can be identified in 
nuclear charge and, if possible, in mass by the correlation of the 
energy versus the energy loss; in the simulation the measured 
nuclear charge is then set equal to the theoretical value, and the 
mass is set equal to the theoretical value up to phosphor and, for 
higher elements, it is deduced from the kinetic energy and the 
time of flight.

	The functioning of the former-generation detector 
\emph{INDRA}~\cite{Pouthas1995,Pouthas1996} is also simulated 
for comparison.
	\emph{FAZIA} differs from \emph{INDRA} because the 
granularity is larger (with a substantial gain in angular 
resolution) and because the mass, in addition to the nuclear 
charge, can be measured.
	When necessary, we will also simulate in the calculation the
existence of neutron detectors imagined as a belt of 
\emph{DEMON}~\cite{Tilquin1995} telescopes covering all azimuthal angles 
and disposed on the longitudinal plane (containing the beam axis).
	For such a configuration, we do not take into account the 
effect of the charged-particle detector on the neutron detection
in the \emph{DEMON} modules.

	In this report, we compare two exclusive experimental 
approaches for reaction experiments with exotic nuclei.
	The first employs \emph{INDRA} (with, if needed , the addition of
\emph{DEMON} modules), where the mass of the fragments (and any isotopic
observable) is not measured, but deduced from the nuclear charge
through the use of a parametrization (EPAX, which supposes that 
the cold nuclei are attracted towards the residue corridor), which
is closely compatible with the decay model GEMINI and which was
adjusted to mostly stable (or moderately exotic) nuclei.
	The second approach is \emph{FAZIA} (with, eventually, the addition 
of \emph{DEMON} modules), which delivers a direct measurement of the
masses, independently of the exoticity of the system.
	Hence the comparison with the response of \emph{INDRA} is useful to appreciate
up to which extent, for the experiment and the observables under examination,
 fragment masses may deviate from the residue corridor and,
once measured by \emph{FAZIA}, bring information on the isovector term
of the nuclear interaction.

\new{
\subsection{Statistics}
	The number of events considered in the transport calculations are 1000. 
	Each one of these events is then used to produce ten decay
paths. 
	For the purpose of simulating the experiment, since the geometrical 
efficiency in the selection of complete events could impose a severe 
reduction of the statistics, the number of events is then largely 
increased by considering rotations around the beam axis.
	At the end, the simulated values, within the uncertainties, correspond 
to one standard deviation of the mean, calculated for a statistics 
of 10000 events (1000 transport events multiplied by 10 decay paths) 
and considering the full error propagation.
 	We would like to stress that the simulated results  
we give should 
be considered as a lower-limit expectation with respect to a real 
experiment.
	The reason is that in a real experiment we can profit from advanced 
tools of event selection and develop the data analysis down to fine 
details, while in the simulation the reduced statistics allows to 
simulate the data analysis only up to a limited degree of accuracy.
	This limitation also propagates to the error bars of the simulated 
results.
}

\new{
	For this same reason, 
we did not exploit the full set of simulations. In fact, in some
cases (especially for the neutron-poor systems at the lower beam energy) the
uncertainties were too large, as compared to the different predictions of the
two asy-EOS.
We stress again that this
would not correspond well to the experimental situation, which is 
already successfully progressing on the study of isospin effects
with stable beams.
	Moreover, for the purpose of comparison with experimental data, 
the accuracy of the simulated results can be improved by increasing
the number of SMF events considered. 
}

\section{
    Simulations: from the selection of the impact parameter to the isotopic observables \label{section2}
}
%
%
\begin{figure}[b]\begin{center}
\includegraphics[angle=0, width=0.9\columnwidth]{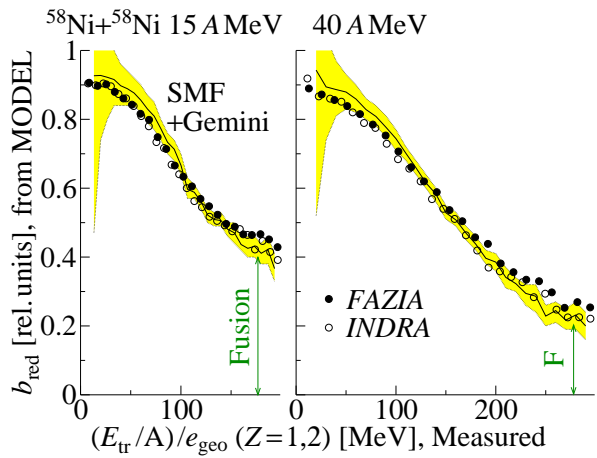}
\end{center}\caption
{
	(Color online)
	Correlation between the impact parameter and the transverse
energy per nucleon for $^{58}$Ni$+^{58}$Ni at 15 and 40 $A$MeV, as
simulated by SMF+GEMINI (solid lines).
	For comparison, the same correlation is simulated for the two 
experimental approaches \emph{INDRA} and \emph{FAZIA}, where the transverse
energy per nucleon is further divided by the geometric efficiency 
$\egeo$.
	The interval in impact parameter which corresponds to fusion is 
indicated.
	The statistical uncertainties of the calculation SMF+GEMINI (asy-soft form)
are indicated by the coloured bands; they are reflected in the 
filtered data with comparable magnitude (Not indicated in the 
figure, for better visibility)
}
\label{fig4}
\end{figure}
%
%
	Experimentally, the impact parameter can be deduced from the
transverse energy per nucleon $\Etr$ of light particles with $Z<3$
(it should be precised that in the SMF simulations no particles with
$Z=2$ are produced; they are generated in the decay, by GEMINI).
	The correlation between the impact parameter and the transverse
energy per nucleon is illustrated in fig.~\ref{fig4} for 
$^{58}$Ni$+^{58}$Ni at 15 and 40 $A$MeV.
	At 40 $A$MeV, the correlation 
extends over all impact parameters except for
$\bred<0.2$, where fusion becomes the dominant mechanism (see fig.~\ref{fig2}).
	At 15 $A$MeV, the correlation is still valid for peripheral
collisions, but the sensitivity is gradually lost when entering the
fusion regime ($\bred<0.55$) and the dissipation becomes total for
complete fusion (around $\bred<0.3$).
	In fig.~\ref{fig4} the measurement of such correlation is
simulated for the two detector arrays \emph{INDRA} and \emph{FAZIA}: the
measured transverse energy per nucleon is divided by the geometric
efficiency $\egeo$ (\emph{INDRA}: $\egeo=0.88$, \emph{FAZIA}: $\egeo=0.78$).
	Both \emph{INDRA} and \emph{FAZIA} perform equivalently well. 
	We shall emphasise that this implies that this same conclusion 
extends to any assembly where a part of one array is replaced by 
the corresponding part of the other, without reducing the whole 
angular coverage.
%
%
\begin{figure}[t]\begin{center}
\includegraphics[angle=0, width=1\columnwidth]{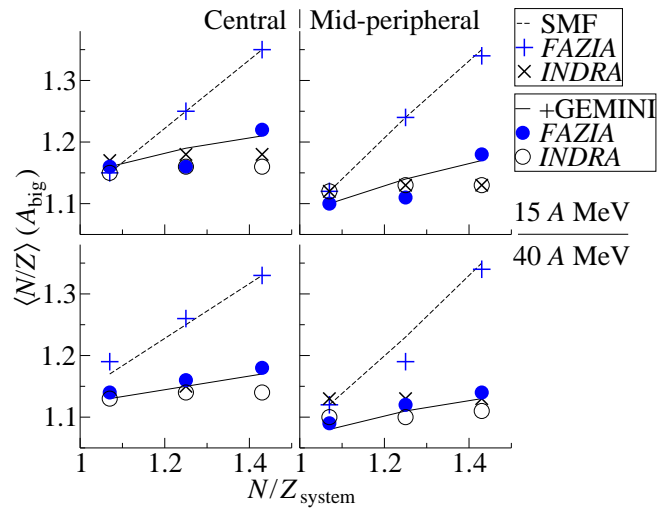}
\end{center}\caption
{
	(Color online)
	Isotopic composition of the largest fragments as a function of
the isotopic composition of the system at 15 (upper panels) and 
40 $A$MeV (lower panels), under the constraint of selecting either
central (left panels) or mid-peripheral events (right panels).
	This observable is simulated for the hot system 
(SMF - asy soft) and for the cold system (SMF+GEMINI).
	For comparison, the same correlation is simulated for the two 
experimental approaches \emph{INDRA} and \emph{FAZIA}.
}
\label{fig5}
\end{figure}
%
%
%

	Profiting of this selection, we focus on the study of the
isotopic composition of the largest fragments.
	This study, presented in fig.~\ref{fig5}, is one example 
intended to answer the question whether the advantage of disposing 
of exotic systems could counterbalance the effect of the 
evaporation process.
	In fig.~\ref{fig5}, three systems $^{58}$Ni$+^{58}$Ni, 
$^{58}$Ni$+^{68}$Ni and $^{68}$Ni$+^{68}$Ni at two incident
energies are studied; to keep an approximate link with the 
underlying reaction mechanism (see fig.~\ref{fig2}), central events 
are defined as $\bred<0.2$ ($\bred<0.55$) for the incident energy 
of $40$ ($15$) $A$MeV; 
mid-peripheral events are defined as $0.25<\bred<0.6$ 
($0.6<\bred<0.75$) for the incident energy of $40$ ($15$) $A$MeV.%
The process of evaporation (GEMINI)
which affects the hot systems (SMF) is more relevant at higher
incident energy.
	Thus, 
larger variations of the isotopic composition ($N/Z$)
of the largest residue ($A_{\mathrm big}$) as a function of the isotopic
composition of the system manifest 
at 15 $A$MeV. 
	Experimentally, \emph{FAZIA} would measure such trend correctly.
Concerning \emph{INDRA}, the redundancy between the parametrisation
used to deduce the mass number from the atomic number (residue
corridor), and the description of the decay by GEMINI globally
results in an apparently good performance for stable systems; for
such systems the evaporation process ends in populating the residue
corridor.
	However, such effect depends on the model. 
	In particular, for the most neutron-rich system,
the mass number parametrisation reveals to be no more sufficient and a direct
measurement of the masses is necessary~\cite{Stoltz2004,Benlliure2008}.
	Such parametrisation will be even less adequate for exotic
systems (which will be studied in future installations exploiting
exotic beams), when the residue corridor is not reached.
	To give an extreme example, we applied the experimental filters
also to the hot systems (SMF), neglecting the decay.
	\emph{FAZIA} measures correctly even these very neutron rich systems,
while \emph{INDRA} associates the residue-corridor masses to the atomic
numbers and gives almost the same result which was previously
simulated for the cold systems (SMF+GEMINI).

	In conclusion, in most studies of isospin effects on the
equation of state, or of thermodynamical
properties in the decay, exotic systems are highly desirable; otherwise, the
smearing effect of the evaporation process will be difficult to surmount.
\section{
    The equation of state: three experimental approaches \label{section3}
}
	The observables studied in fig.~\ref{fig5} led to the
conclusion that secondary decay may impose major difficulties in 
the experimental study of the equation of state.
	Exotic nuclei may be advocated as more suited systems, where
the secondary decay may be better retraced as far as the residue
corridor is not reached in the cooling process.
	Alternatively, we can also search for other observables which are
less sensitive to the secondary decay.
	Three possibilities are explored in this section: the 
imbalance ratio, first introduced by the MSU 
group~\cite{Tsang2004,Chen2005}; the high-energy part of neutron 
and proton spectra, proposed in the framework of BUU~\cite{Li2006}, 
and also successfully exploited via IQMD~\cite{Zhang2008}; and
finally, the neck emission in ternary events, studied at length by 
the Catania group~\cite{Baran2005,DeFilippo2005}.

\subsection{
	Imbalance ratios}
%
%
\begin{figure}[b]\begin{center}
\includegraphics[angle=0, width=0.8\columnwidth]{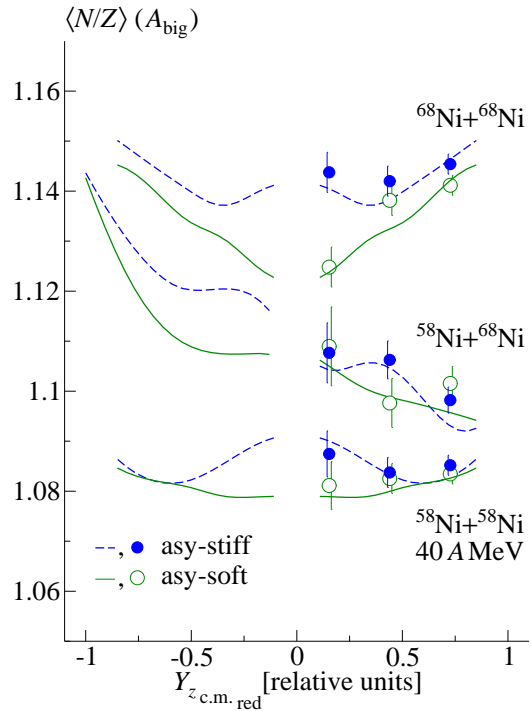}
\end{center}\caption
{
	(Color online)
	Same observable studied in fig.\ref{fig3}; we focus on the cold systems (SMF+GEMINI) of the 
system at 40 $A$MeV and we add the filtered data, simulated for the experimental device 
\emph{FAZIA}.
}
\label{fig6}
\end{figure}
	The isospin-transport ratio, introduced by 
F.Rami~\cite{Rami2000} to study isospin equilibration, has been 
exploited by the MSU group~\cite{Tsang2004,Chen2005,Tsang2009}
because of its sensitivity to the density dependence of the 
symmetry energy.
	In the case of the Ni isotopes of the present study, the 
imbalance ratio is defined as
\begin{equation}
	R=\frac{2x_{58+68}-x_{68+68}-x_{58+58}}{x_{68+68}-x_{58+58}},
\label{inbalance}
\end{equation}
where $x$ is an isospin sensitive observable, possibly linearly 
correlated with asymmetry, measured in the three different 
reactions.
	The motivation of introducing such observable is that, by 
focusing on the differences in isospin observables between mixed 
and symmetric systems, $R$ is expected to largely remove the 
sensitivity to preequilibrium emission and enhance the sensitivity 
to isospin diffusion between projectile and target.
%
%
\begin{figure}[t]\begin{center}
\includegraphics[angle=0, width=0.9\columnwidth]{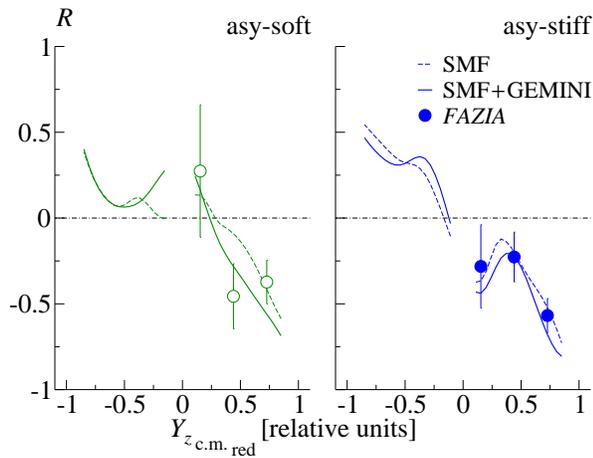}
\end{center}\caption
{
	(Color online)
	Imbalance ratio (see text) for the asy-stiff and asy-soft 
parametrisations, corresponding to the same observable studied 
in fig.\ref{fig3}, for the systems $^{58}$Ni$+^{58}$Ni $^{58}$Ni$+^{68}$Ni and 
$^{68}$Ni$+^{68}$Ni at 40 $A$MeV, for all impact parameters, and as
a function of the longitudinal component of the rapidity in the 
centre of mass relative to the projectile (target) rapidity for positive (negative) rapidities.
	The statistical uncertainties of the model are indicated 
by the error bars.
}
\label{fig7}
\end{figure}
	We have undertaken this analysis for the reactions at 40 AMeV, fow which the
statistics of binary events is good enough.  

	Considering all impact parameters, we assigned the average 
isospin content of the largest fragment, not produced in fusion events,
to the $x$ variable. 
	This observable, explored in fig.\ref{fig3}, is shown again in fig.\ref{fig6},
together with the corresponding response of the experimental device
\emph{FAZIA}. From a closer inspection of figs.\ref{fig3},\ref{fig6}, 
one may notice that the effect of the secondary decay on this
observable is not the same in the different rapidity bins. 
	This can be ascribed to the fact that the excitation energy of 
primary fragments is not constant, but it is larger at mid-rapidities, 
which correspond to more central events. 
	However, in each rapidity bin, this observable is expected to be correlated
to its value prior to decay.
	From these data points fig.\ref{fig7} is constructed.
	We may remark that, since the isotopic composition of the 
asymmetric system is intermediate between the symmetric systems,
according to the definition of Eq.\ref{inbalance}, the imbalance ratio should approach zero in the 
vicinity of midrapidity and evolve towards negative values in the 
projectile side for increasing rapidity and eventually reach the 
value of $-1$ for large transparency; in the target side, the 
imbalance ratio should evolve for increasing negative rapidity
towards the same values explored in the projectile side with 
opposite sign.
	This overall behaviour, reflected by the hot fragments (SMF), 
is well reproduced also after the evaporation process (SMF+GEMINI);
	this shows that, indeed, the effects of secondary decay can be 
removed when suitable observables which combine differences and 
ratios of isospin dependent properties are adopted, such as the 
imbalance ratio.
	The difference between the predictions of the asy-stiff and asy-soft 
parametrisations is preserved after the secondary decay stage and is measurable
with FAZIA.
	Comparing fig.~\ref{fig6} and fig.~\ref{fig7}, we can see 
that the discrimination between the two parametrisations is 
essentially due to the presence of the neutron rich system $^{68}$Ni$+^{68}$Ni, 
which is the only case where the residue corridor is not reached 
during the deexcitation. 
	Following the discussion of fig.~\ref{fig5}, we expect that 
the imbalance ratio will provide a better discrimination between 
different effective interactions with the use of more exotic beams.

	A similar analysis could be performed taking the transverse 
energy $\Etr$ of light particles as a centrality selector (see 
fig.~\ref{fig4}) instead of the rapidity of the largest fragments. 
	However, the latter appears more directly connected to the 
dissipation mechanism between the two reaction partners, while 
$\Etr$ may be influenced by additional effects, such as 
pre-equilibrium dynamics and cluster-emission mechanisms. 
	For this reason, the analysis proposed in fig.~\ref{fig7} 
should be more helpful when comparing experimental data with the 
predictions of transport models which differ by the treatment of 
correlations~\cite{Rizzo2008}.
	Also different isospin-sensitive observables may be used, such 
as the isoscaling parameter and the isobaric yield ratio 
$\ln(Y(^{7}\mathrm{Li})/Y(^{7}\mathrm{Be}))$ employed by the MSU 
group~\cite{Tsang2009}, or the average $N/Z$ of the 
light-charge-particle emission~\cite{Galichet2009}.
	This would be possible within our analysis protocol, but it
demands a larger statistics for the secondary-decay treatment.
	For all those choices, as far as the observables are linearly 
correlated to the N/Z of projectile-like and target-like 
fragments~\cite{Tsang2009,Galichet2009}, the corresponding
results for the imbalance ratio should be rather close to the ones 
depicted in fig.~\ref{fig7} for the observable we adopted.

\subsection{
    Light particles: neutron and proton spectra}
	The evolution of isotopic ratios of light particles as a
function of their kinetic energy and the isotopic composition of 
the system $(N/Z)_{\textrm{system}}$ has also been shown to be 
sensitive to the stiffness of the equation of 
state~\cite{Li2006,Zhang2008}.
	When light particles (like $t$ and $^3$He) are concerned, the
SMF model can no more access such observable which, in general, is 
strongly dependent on the model.
	For these reasons, in this section we restrict to the study of 
preequilibrium neutrons and protons.

%
%
\begin{figure}[b]\begin{center}
\includegraphics[angle=0, width=0.9\columnwidth]{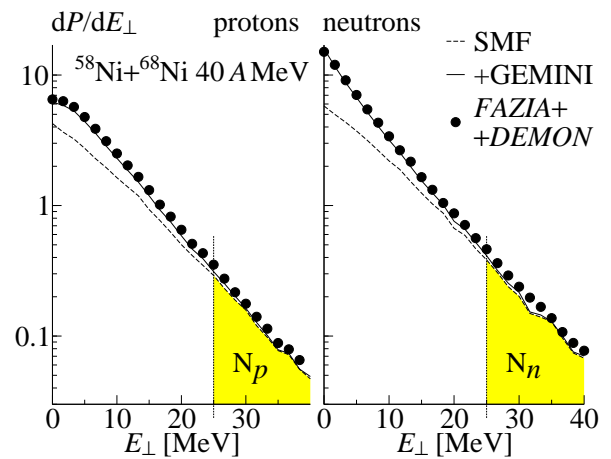}
\end{center}\caption
{
	(Color online)
	Proton (left panel) and neutron (right panel) yields as a 
function of $E_{\perp}$ for the system $^{58}$Ni$+^{68}$Ni at 40 
$A$MeV.
	\new{Units are in percent of the total neutron yields.}
	The spectra are simulated for the hot system (SMF - asy-soft) 
and for the cold system (SMF+GEMINI).
	The same spectra are simulated as measured with an experimental
device composed of \emph{FAZIA} coupled with \emph{DEMON} modules.
}
\label{fig8}
\end{figure}

	The neutron and proton yields as a function of the
perpendicular component of the kinetic energy $E_{\perp}$, are illustrated in 
fig.~\ref{fig8} for the system $^{58}$Ni$+^{68}$Ni at 40 $A$MeV.
	The spectra are simulated before (SMF) and after (SMF+GEMINI)
the secondary decay.
	As expected, the low-energy side of the spectra is mostly affected by the
secondary decay.
	To minimise the effect of secondary decay, we select the
high-energy side of the spectra where the two simulations, before
and after evaporation, coincide: we impose $E_{\perp}>25$MeV.

	The shape of the proton spectra can be measured directly by
an experimental device like \emph{FAZIA}, as simulated in
fig.~\ref{fig8}, or like \emph{INDRA}.
	The shape of the neutron spectra can  be deduced from
the proton spectra only if complete events can be recorded; 
since this is practically impossible to achieve at intermediate energies 
due to the high multiplicity of the events, 
the neutron spectrum should be measured directly with the use of a neutron
detector; in the present simulation, shown in fig.~\ref{fig8}, we suppose that 
the detector \emph{DEMON} is employed.
	In particular, we suppose to dispose
of \emph{DEMON} telescopes covering all azimuthal angles, and only an
incomplete interval of 30 deg in zenith angle, so that the shape of
neutron spectra can be measured correctly.
	For such a configuration, we do not take into account the 
effect of the charged-particle detector on the neutron detection, 
even if, in a realistic configuration, such effect could be large 
and the neutron detectors may have to be even substituted to some 
charged-particle detectors.
	It is possible, but it should be proved experimentally, that analogous 
information may be deduced from the analysis of $t$-$^3$He spectra, 
which would make this analysis accessible to neutron-blind devices 
like \emph{INDRA} or \emph{FAZIA} alone.
	The integral of the proton spectra can be measured directly, by
selecting only complete events.
	The integral of the neutron spectra can not be measured
directly in any case: in general, the efficiency of neutron
detectors would not be sufficient to impose a condition of event
completeness even when using a $4\pi$-array.
	The integral of neutron spectra is deduced by difference, from
the measurement of the nuclear charge and mass of fragments and
particles detected in complete events.
	Hence, for  a correct evaluation of the efficiency of the neutron detector, 
also the mass detection, as provided by \emph{FAZIA}, is necessary.
	The slight discrepancy between the theoretical and measured
spectra in the simulation of fig.~\ref{fig8}, both for neutron and
proton spectra, is due to the choice of defining as complete events
those events where less than 15\% of the total nuclear charge is
missing.
%
%
\begin{figure}[b]\begin{center}
\includegraphics[angle=0, width=1.\columnwidth]{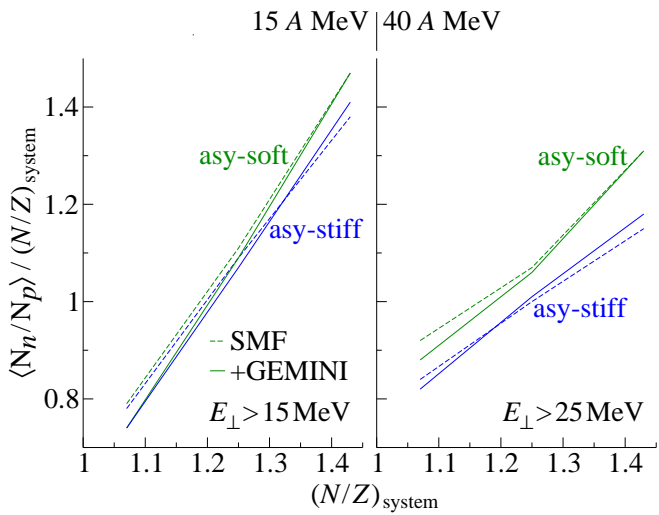}
\end{center}\caption
{
	(Color online)
	Ratio between the number of neutrons over the number of protons
averaged over complete events and normalised to the isotopic 
composition of the system, as a function 
of the isotopic composition of the system.
	This observable is simulated for incident energies of 15 and 40
$A$MeV, for hot systems, before the secondary decay occurs (SMF) 
and for cold systems, after the secondary decay (SMF+GEMINI).
	The two prescriptions for the equation of state, asy-stiff and 
asy-soft are employed.
	Specific event selections are imposed depending on the incident
energy. 
	15 $A$MeV: only neutrons and protons with $\Eperp>15$ MeV are chosen.
	40 $A$MeV: only neutrons and protons with $\Eperp>25$ MeV are chosen.
	In both cases, only events not leading to fusion are selected.
}
\label{fig9}
\end{figure}

	From the neutron and proton spectra we deduce the evolution of
the ratio between the number of neutrons over the number of protons
$\textrm{N}_n/\textrm{N}_p$, averaged over complete events and normalised to the
isotopic composition of the system $(N/Z)_{\textrm{system}}$, as a
function of the isotopic composition of the system.
	In fig.~\ref{fig9} this observable is studied for hot systems,
before and after the secondary decay, for the two incident energies 15 and 40
MeV per nucleon. 
	We mention that the condition  $E_{\perp}>15~$MeV is used 
for the reactions at 15 AMeV, due to the lower energy available in 
this case for the pre-equilibrium emission.
	It has been already observed that the best sensitivity to 
the equation of state is obtained in central collisions~\cite{Li2006}.
	However, especially at 15 MeV per nucleon, 
it is difficult to extract central events from the $\Etr$ distribution, see
fig.~\ref{fig4}. Hence, 
	for both incident energies, we consider all impact parameters,
but we select fusion events out.
	The different parametrisations for the equation of state,
asy-stiff and asy-soft, do not determine any appreciable
difference at 15 MeV per nucleon.
	This could be due to the fact that, at this low energy, compression-expansion
effects are quite reduced and, while pre-equiquilibrium emission takes place, 
the nuclear density keeps rather close to the saturation value, where
the two parametrisations give, by construction, the same value of the symmetry
energy.
	On the other hand, at 40 MeV per nucleon, the two 
parametrisations of the symmetry energy lead to clearly distinct 
behaviours even for non-exotic systems: the normalised average 
neutron-to-proton ratio is up to 13\% larger for the asy-soft case.
	Now the isotopic content of the pre-equilibrium emission is sensitive
to the low-density behaviour of the symmetry energy, since particles mostly
escape while the composite nuclear system is expanding.
\new{ 
	A neutron-richer emission is seen in the asy-soft case, corresponding to the higher value of the symmetry energy below normal density (see fig.~\ref{fig1}).
}
One can also notice that the dependence of this observable on the system initial
asymmetry is flatter for the reactions at 40 AMeV. This reflects the higher energy
available, with respect to the 15 AMeV case, at the pre-equilibrium stage.
	In the following, we concentrate on the systems at 40 MeV per
nucleon.
%
%
\begin{figure}[t]\begin{center}
\includegraphics[angle=0, width=0.95\columnwidth]{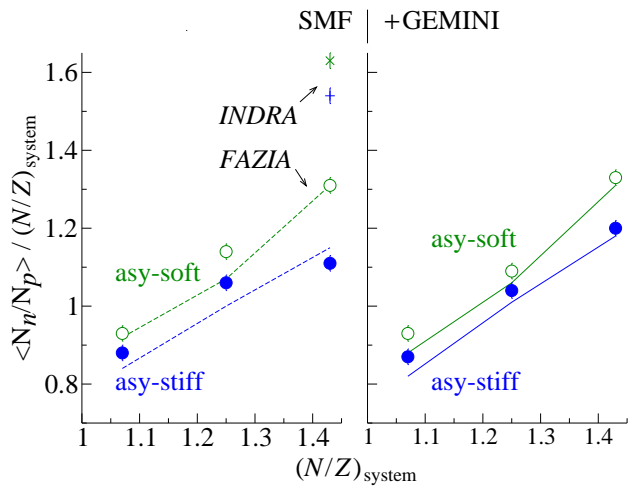}
\end{center}\caption
{
	(Color online)
	Same observable as in fig.~\ref{fig9}, for an incident energy 
of 40 $A$MeV. 
	Both hot systems (SMF, left panel) and cold systems 
(SMF+GEMINI, right panel) are simulated as measured with an 
experimental device composed of \emph{FAZIA} coupled with \emph{DEMON} modules
before and after secondary decay.
	The simulation of \emph{INDRA} coupled with \emph{DEMON} modules
is added for the most neutron-rich system, before secondary decay.
}
\label{fig10}
\end{figure}
%
%
%
       We observe that the signals are still 
preserved after the secondary decay, thanks to the energy selection 
we adopted.
	In fig.~\ref{fig10}, we simulated the response of the two detection
devices, \emph{FAZIA} and \emph{INDRA}.
	The error bars account for the accuracy in the correction for the
efficiency of the neutron detection, supposing that it is known with an 
uncertainty of around 5\%; additional systematical uncertainties
(not shown) would be produced by the procedure of accounting for 
the number of fragments and particles which are not detected.

	The simulation demonstrates that \emph{FAZIA} would perform
 efficiently and succeed in measuring the signal.
%
	However, after the deexcitation stage, the mass of the evaporation residues,
which should be measured in order to reconstruct the number of emitted neutrons,
approaches the residue corridor.
	In order to study a more constraining situation, for testing 
purposes, we replace the cold system by the hot system, before the decay, 
in the experimental filter.
	For the most neutron-rich system, we add simulated points representing INDRA,
which does not measure the masses, but deduces them through the evaporation-corridor
prescription introduced in the detector simulation: these points, 
as they are referred to a very neutron-rich system, far
from the residue corridor, result in fact to be incompatible.
	We conclude that such signal survives after the secondary decay 
and is measurable only with a device like \emph{FAZIA} in conditions 
	where the   reaction products are far from the residue corridor.

\subsection{
    Neck emission in ternary events}
	It has been discussed in the literature~\cite{Baran2005} 
that the dissipative dynamics leading to neck formation in 
mid-peripheral collisions at Fermi energies is strongly sensitive 
to the isospin transport properties and to the isovector properties 
of the equation of state.

	We focus on the system $^{68}$Ni$+^{68}$Ni at 40 $A$MeV; 
such incident energy is large enough to exceed the threshold for 
the formation of neck fragments.
	For this specific simulation the stopping time was fixed to
160 fm/c, which is long enough to follow the dynamics of the
neck formation.
	We select ternary events where, in addition to light particles,
we observe only the following three fragments: one quasi-projectile
residue, one quasi-target residue, and one intermediate-mass
fragment at midrapidity, which corresponds to the neck fragment.
	In addition, we impose two more constraints for the selection of
ternary events.
	First, we reduce to semiperipheral collisions where $\bred$
varies in the interval $0.45<\bred<0.75$: in fig.~\ref{fig2} we
already signalled the presence of neck events in this interval. 	
	The second constraint is on the mass number of the third
fragment: by adjusting a selection threshold to $Z\ge 5$, we 
ensure that the ternary events with three IMF's that we select can
not be confused with secondary-decay events which produce fragments
of small size.

 	The observable we investigate is the isotopic composition of
the neck fragment in ternary events, selected as defined above.
	This system, studied before the secondary decay, is less neutron
rich when the symmetry energy is parametrised with the asy-soft 
term, and more neutron rich when an asy-stiff interaction is used.
	This is due to isospin migration effects from projectile and target
towards the low-density region of the neck. 
\new{
	In particular, the evolution of the isospin asymmetry of the 
neck region is conditioned by the contribution of the migration 
process to the currents of neutrons and protons, driven by density 
gradients.
	Such contribution is represented in eq.~\ref{eq_3} by the term 
proportional to the derivative of the symmetry energy and the 
density gradient.
	The process of migration determines an enrichment in the 
neutron content of the diluted neck region, which is larger for 
larger variations of the symmetry energy as a function of the 
density, and is therefore more effective for a stiffer form of the 
EOS around normal density , which is in this case given by the 
asy-stiff parametrisation (see fig.~\ref{fig1}).
}
	When the secondary-decay process is added up, in the asy-soft
case, the more excited neck fragments end up in the evaporation corridor and
their $N/Z$ ratio reduces.
	In the asy-stiff case, the hot system is more neutron-rich and
the decay path does not reach the evaporation corridor.

%
%
\begin{figure}[b]\begin{center}
\includegraphics[angle=0, width=0.95\columnwidth]{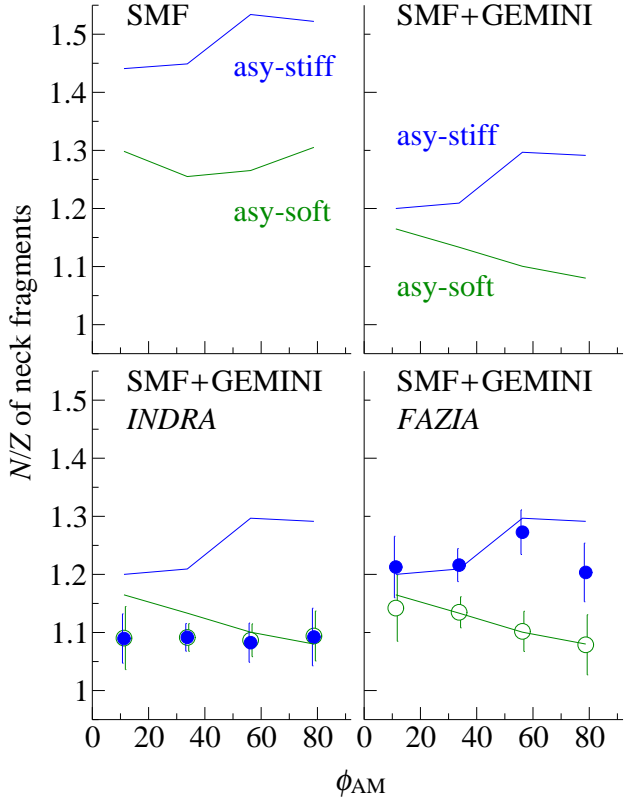}
\end{center}\caption
{
	(Color online)
	Isotopic composition of the neck fragment in ternary events, 
selected as defined in the text, as a function of $\phiAM$, for 
the system $^{68}$Ni$+^{68}$Ni at 40 $A$MeV.
	The spectra are simulated for the hot system (SMF, upper left panel) and for the 
cold system (SMF+GEMINI, upper right panel).
	The spectra representing the cold system are simulated as 
for the two experimental approaches \emph{INDRA} (lower left panel) and \emph{FAZIA} (lower right panel).
}
\label{fig11}
\end{figure}
	Such difference in shown in fig.~\ref{fig11}, where the
isotopic composition of neck fragments in ternary events is plotted
as a function of $\phiAM$, where $\phiAM$ is the smallest angle 
which measures the angular misalignment between the velocity vector 
of the neck fragment in the centre of mass frame and the direction 
along which the two heaviest fragments (representing approximately 
the quasi-projectile and the quasi-target) are aligned.
	This convention, followed in this paper, does not correspond exactly
to the definition of $\phiplane$~\cite{DeFilippo2009}, more often 
followed in experimental studies of the phenomenon of nonequilibrium 
fission.
	We signal that such studies are closely compatible with the 
present work and explicitly oriented to the investigation of the EOS.

	The increasing trend observed in the asy-stiff case may be due
to the impact parameter mixing in our analysis. Indeed, fragments with
larger transverse velocities are more easily emitted in 
the most central events of the selected interval of impact parameters, 
for which, due to the longer reaction time, 
isospin migration is more effective. The trend is less clear in the
asy-soft case, where isospin migration effects are less pronounced.
For the same reason, fragments having  $\phiAM$ close to 90 deg
are more excited.
	In the asy-stiff case, the isotopic composition of neck
fragments reduces almost systematically when passing from the system
before decay (SMF) and after decay (SMF+GEMINI).
{The effects related to the different excitation energy are more evident
in the asy-soft case:}
the less excited fragments experience a short decay path and
are close to the isotopic composition of the hot fragments, while
the fragments which could spend more excitation energy in the decay
reached the evaporation corridor.
	The further addition of the experimental filter demonstrates
that \emph{FAZIA} would measure this observable precisely.
	On the other hand, \emph{INDRA} would not succeed in measuring this
observable, and the discrepancy is evident when the residue
corridor is not reached, that corresponds to the asy-stiff case.
	In conclusion, the asymmetry of the neck fragments appears as a
promising observable to be exploited in the investigation of the
behaviour of the symmetry energy below saturation density.
Due to the isospin-migration mechanism, these fragments are particularly
neutron-rich, especially in the asy-stiff case, and the effect survives
to the secondary-decay stage. 

\new{Isospin migration is more effective in neutron-rich systems, 
however interesting results have been reported also
in the case of stable, less asymmetric systems, in theoretical \cite{lionti}, 
as well as in experimental analyses \cite{DeFilippo2009}.}
It would be extremely interesting to pursue
the investigation of the neck dynamics also at lower beam energy.
Though the analysis is complicated by the reduced statistics of ternary
events, fragments are expected to be less excited, opening
the possibility to access more directly the genuine dynamical effects.
Work is in progress in this direction.

\section{
    Conclusions}
    In summary, we built a protocol for simulating reaction experiments 
which profit from next-generation radioactive-beam facilities and 
from innovative experimental solutions; these latter are 
specifically dedicated to the measurement of isotopic observables 
in exclusive experiments at low and intermediate energies.
    In this work, the systems 
$^{68}$Ni$+^{68}$Ni, $^{58}$Ni$+^{68}$Ni and $^{58}$Ni$+^{58}$Ni 
at 15 and 40 MeV per nucleon were simulated for
a continuous distribution of impact parameters $b$ within the 
Stochastic Mean Field Model, with the addition of a secondary-decay 
process (described within the model GEMINI).
    The experimental conditions were simulated by supposing that the 
experimental observables are delivered either by the 
former-generation detector \emph{INDRA}, or by a next-generation 
detector, which corresponds to the \emph{FAZIA} project; 
such project, under development, is intended to build a 
$4\pi$-array of silicon detectors which, beside measuring the 
nuclear charge of the fragments, are also sensitive to the mass and 
have low thresholds.
    Such simulation protocol was used to discuss new experimental 
solutions for probing the nuclear equation of state in heavy-ion 
collisions at Fermi energies.

    Firstly, we studied the effect of the secondary decay. 
    Such process, which creates a significant bias in the 
determination of the EOS properties, should be taken into account 
in the choice of the experimental strategy.
	According to our simulations, secondary-decay effects can be 
partially cured by studying exotic systems. 
	However, even with exotic beams, secondary decay has to be 
controlled before a reliable extraction of the isovector equation 
of state can be obtained.
   
	The strategy of constructing suitable combinations of isotopic 
observables (such as imbalance ratios), that was proposed for 
similarly prepared systems, appears to be  an efficient way to 
reduce the effects of the secondary decay.
	Other  possible solutions consist in selecting observables 
connected to fragments which are formed with insufficient 
excitation for feeding any relevant decay process, or uniquely 
connected to the initial phases of the reaction process, like in 
the preequilibrium stage.

    Following this second approach, we identified two more probes 
which are negligibly affected by the secondary decay: the relative 
yields of preequilibrium high-energy light particles (we focused on 
neutrons and protons), and the isotopic composition of neck 
fragments in ternary events.
    Both observables would exhibit significantly different 
signatures for the asy-stiff or the asy-soft behaviour of the 
symmetry energy, even when the secondary decay is accounted for.
\new{
	This is particularly evident in the case of the neutron-rich
$^{68}$Ni~$+^{68}$Ni system, for which isospin effects are 
enhanced.}   

\new{
	In this framework, observables related to the neck dynamics
appear rather promising. 
	Interesting effects have been revealed already with stable 
beams \cite{DeFilippo2009,Amorini2009}, which may constitute one 
	passage towards exotic-beam experiments. 
	Moreover, it would be very interesting to extend this analysis 
to lower beam energies.} 
    We found that these signatures could be identified with 
an exclusive detector device which is compatible with the \emph{FAZIA} 
project, and which is based on the simultaneous detection of 
charges and masses over a large solid angle.
	The principle of such an experimental approach has several 
advantages.
	First of all, the possibility of performing a complete event 
reconstruction through the data analysis allows to refine the 
criteria of centrality selection, which is a crucial point in the 
comparison between theoretical simulations and experimental data.  
	Then, the measurement of the mass of all reaction products 
gives access to more sophisticated isospin-sensitive observables.
	Finally, we shall also stress that such experimental approach 
would be suited for analysing within the same experiment several 
probes for the symmetry energy, which are also sensitive to 
different stages of the reaction.

\section{
	Acknowledgements}
	The efficiency diagram used to simulate the \emph{DEMON} neutron 
telescopes was kindly provided by Frank Delaunay through a 
GEANT4 simulation.

%
%
%

\end{document}